No revision, but the original submission corrupted in e-mail process.
One figure is mailed on request in the form of postscript file.
-----------------------------------------------------------------------

\documentstyle[12pt]{article}
\newcommand{\beql}{\begin{eqnarray}}
\newcommand{\eeql}{\end{eqnarray}}
\newcommand{\beq}{\begin{equation}}
\newcommand{\eeq}{\end{equation}}
\newcommand{\bc}{\begin{center}}
\newcommand{\ec}{\end{center}}
\newcommand{\bfr}{\begin{flushright}}
\newcommand{\efr}{\end{flushright}}
\newcommand{\bfl}{\begin{flushleft}}
\newcommand{\efl}{\end{flushleft}}
\newcommand{\bl}{\begin{large}}
\newcommand{\el}{\end{large}}
\newcommand{\bll}{\begin{Large}}
\renewcommand{\ell}{\end{Large}}
\newcommand{\blll}{\begin{LARGE}}
\newcommand{\elll}{\end{LARGE}}
\newcommand{\bdes}{\begin{description}}
\newcommand{\edes}{\end{description}}
\newcommand{\bitem}{\begin{itemize}}
\newcommand{\eitem}{\end{itemize}}

\newcommand{\benum}{\begin{enumerate}}
\newcommand{\eenum}{\end{enumerate}}

\newcommand{\hsi}{\hspace{0.5cm}}
\newcommand{\vs}{\vspace{0.5cm}}
\newcommand{\vsl}{\vspace{1cm}}

\newcommand{\blarge}{\begin{Large}}
\newcommand{\elarge}{\end{Large}}
\def\lromn#1{\uppercase\expandafter{\romannumeral#1}}

\def\blist{\begin{list}{\setlength{\rightmargin}{\leftmargin}}}
\def\elist{\end{list}}
\begin{document}

\begin{titlepage}

\begin{flushright}
\begin{large}
TU/93/440\\
July 1993
\end{large}
\end{flushright}

\vspace{12pt}

\begin{center}
\begin{Large}

\renewcommand{\thefootnote}{\fnsymbol{footnote}}
\bf{Wormhole and Hawking Radiation}\footnote[1]
{Work supported in part by the Grant-in-Aid for Science Research from
the Ministry of \\ \hspace*{0.6cm} Education, Science and Culture of Japan No.
0}

\end{Large}

\vspace{36pt}

\begin{large}
\renewcommand{\thefootnote}{\fnsymbol{footnote}}
M.Hotta
and M.Yoshimura\footnote[2]
{E-mail address: YOSHIM at JPNTUVM0.\ BITNET\@.}\\
Department of Physics, Tohoku University\\
Sendai 980 Japan\\

\vspace{72pt}

{\bf Abstract}\\

\end{large}
\end{center}

\vsl
It is shown in a variant of two dimensional dilaton gravity theories
that an arbitrary, localized massive source put
in an initially regular spacetime
gives rise to formation of the wormhole classically, without
accompanying the curvature singularity. The semiclassical
quantum correction under this wormhole spacetime yields Hawking radiation.
It is expected, with the quantum back reaction added to the classical
equation, that the information loss paradox may be resolved in this
model.

\vspace{12pt}

\end{titlepage}

The end point of the black hole evaporation poses a challenging conceptual
problem that must be solved before one embarks on a full-fledged theory of
quantum gravity. If the black hole is an inevitable consequence of the
gravitational collapse, then one appears to have no alternative choice but
to accept the thermal spectrum of Hawking radiation \cite{hw75}.
The question is then: does a quantum mechanical
pure state  develop into a mixed thermal state ? To answer this question
in a definitive way, one presumably has to
deal with the problem of quantum back reaction to
the background spacetime, in this case,
caused by Hawking radiation. Some interesting toy model in two dimensions
has been proposed recently to discuss this issue at depth,
however with inconclusive results so far with
regard to the end point of the black hole evaporation \cite{cghs},
\cite{rstetc}.

Among others, an interesting alternative
resolution to the clash between gravity
and quantum mechanics has been suggested in Ref.\cite{dyson} and
Ref.\cite{whinf}, which take the view that
the apparent loss of information in the Hawking radiation is actually
recovered by information flow into a disconnected world in the interior of
the black
hole inaccessible to this world. If this is the solution, the curvature
singularity inside the black hole, which normally exists, must be replaced
by something like a wormhole.
Whether and how this might be realized is
not known to the best of our knowledge.

In this paper we would like to provide such a mechanism
related to the wormhole formation and its subsequent evaporation
in a concrete two dimensional model. The model is a variant of two dimensional
dilaton gravity theories considered by Callan, Giddings, Harvey, and
Strominger (CGHS) \cite{cghs}.
A unique classical solution without singularity is found for an arbitrary
localized massive source, and is shown to yield the wormhole geometry
(Lorentzian) with global event horizon.
Moreover, in the large $N$ semiclassical limit the wormhole
background gives rise to Hawking radiation. Although we still lack
a quantitative semiclassical analysis with the back reaction included in
the form of the usual Polyakov term,
it is guaranteed in this model that time evolution does not lead to
the curvature
singularity even when quantum back reaction is incorporated. Thus either of
the two possibilities with regard to the end point of the wormhole evaporation
exists; a complete disappearance leaving behind no horizon,
or evolution to a still unknown quantum
remnant of the wormhole structure remaining behind event horizon.

We consider the two dimensional dilaton gravity defined by the
classical action,
\beql
S_{r} = \frac{1}{2\pi} \int\! d^{2} x \sqrt{-g}\;[\:e^{-2\varphi}
  \:( -R - 4\,\partial_{\mu}\varphi\,\partial^{\mu}\varphi + 4\lambda^{2})
+L^{(m)}\:]\,.
\eeql
This model was called the reversed CGHS model \cite{my92},
because it differs from the original CGHS model \cite{cghs},
in the two signs of the curvature and the dilaton kinetic terms. Despite
the sign reversal this model describes a reasonable toy model of gravity
in the weak field limit \cite{hy93-1}.
The cosmological constant $\lambda^2$ sets a length scale in the theory.
A novel feature of the dilaton gravity is that the coupling factor
$e^{2\varphi}$ acts as a varying gravitational constant,
as in the Brans-Dicke theory in four dimensions.
For the matter part $L^{(m)}$ we take a massive field as the source of
the gravity force.

The field equation in the conformal gauge,
\(\;
ds^{2}=\,-e^{2\rho}dx^{+}dx^{-}
\:\)
with
\(\:
x^{\pm}=x^{0}\pm x^{1}\,,
\:\)
is written as
\beql
\partial_+ \partial_- e^{-2\varphi} -\lambda^2 e^{2\rho -2\varphi}
&=&\pi\, T_{+-}\,,\\
\partial_+ \partial_-(\rho-\varphi)&=&-\,\frac{\pi}{2}
\,e^{2\varphi}\,T_{+-}\,,\\
\partial_{\pm}^2 e^{-2\varphi} -2\,\partial_{\pm}(\rho-\varphi)\,
\partial_{\pm} e^{-2\varphi} &=& -\,\pi\,T_{\pm\pm}\,.
\eeql
The stress tensor $T_{\mu\,\nu}$ for the general localized massive source
consistent with the conservation is given by
\begin{equation}
T_{++} =T_{--} =T_{+-} = -\,\frac{M}{4\pi}\,e^{\rho}\,\delta (x^1 )\,,
\label{source}
\end{equation}
with $M=\pi m$, $m$ a point source mass \cite{hy93-1}.
The massive source is located at $x^1 =0$.

Exact classical solution, along with the scalar curvature $R$,
is then given by
\begin{eqnarray}
e^{-2\varphi}&=& e^{-2\lambda x^0 }
\left(A-\frac{M}{2\lambda B} \sqrt{1+B e^{2\lambda (x^0 - |x^1 | )}\,}
\right)+AB,\\
e^{2\rho}&=&e^{2\varphi} e^{-2\lambda x^0 }
\left(A-\frac{M}{2\lambda B}\frac{1}{\sqrt{1+B e^{2\lambda(x^0 - |x^1 | )}\,}}
\right),\\
R&=&\frac{2Me^{2\lambda x^{0}}}{A\sqrt{1+Be^{2\lambda x^{0}}\,}
-\frac{M}{2\lambda B}}\,\delta\,(x^{1})-\frac{4\lambda^{2}AB
\,e^{2\lambda x^{0}}}{A-\frac{M}{2\lambda B}\sqrt{1+B e^{2\lambda
(x^0 - |x^1 | )}\,}+ABe^{2\lambda x^{0}}}\,.
\end{eqnarray}
The parameters, $A\,$ and $B\,,$ are integration constants.
This set of solutions exhausts all classical solutions with
the reflection symmetry under $x^1 \rightarrow -\,x^1 $ when the localized
source is given by Eq.\ref{source}, as will be
shown in a subsequent paper \cite{hy93-3}.
The divergent curvature at the source $\propto \delta\,(x^1)$
should be regarded as an artifact of
the localized matter distribution. With an extended source this component
of the curvature is expected to be smeared out.

To uncover the spacetime structure of these classical solutions,
it is important to extend the coordinate maximally. For this purpose,
we introduce a new coordinate system according to
\(\:
\tilde{x}^{\pm}=-\,(\,1/\lambda)\,e^{-\lambda x^{\pm}}\,,
\:\)
for $x^{1}>0$.
To obtain the form of solution in $x^{1}<0$, we replace the coordinates
by the rule,
$\tilde{x}^{+}\leftrightarrow\tilde{x}^{-}$.
The range of the new coordinates, $\tilde{x}^{\pm}$,
is then extended to the whole real axis.

Unless $A=0$, all these solutions have both time-like curvature and dilaton
singularity at the same location,
\beql
\tilde{x}^{+}=-\,\frac{AB}{\lambda^{2}\,A\tilde{x}^{-}
+\frac{M}{2B}\,\sqrt{(\lambda\tilde{x}^{-})^{2}+B\,}}\,,
\eeql
for $\tilde{x}^+ <\tilde{x}^-$, and at a similar location in the other region,
$\tilde{x}^+ >\tilde{x}^-$.
We reject these singular solutions as a model of the gravitational collapse,
since one cannot set up a regular boundary condition at the null past
infinity for these solutions.

When $A=0$, two types of solutions with different signs of $B$ exist, but
are related
to each other by a coordinate transformation. By a suitable choice of
the origin of time coordinate $x^0$, one may set $B=-\,\frac{M}{2\lambda}$ and
gets the unique solution without singularity,
\begin{eqnarray}
e^{-2\varphi}& =& e^{-2\lambda x^0 }
\sqrt{1-\frac{M}{2\lambda}\,e^{2\lambda (x^0 - |x^1 | )}\:},\\
e^{2\rho}&=&
\frac{1}{1-\frac{M}{2\lambda}
\, e^{2\lambda (x^0 - |x^1 | )}\,},\label{eqn101}\\
R&=&2Me^{2\lambda x^0} \delta (x^1 )\,.
\end{eqnarray}
It is not necessary to extend the coordinate patch in this case.

The spacetime described by this metric is everywhere flat, except at
the location of the source. To see this and detailed spacetime structure
more clearly, it is useful to introduce the
asymptotically flat coordinate with
\( \:
ds^2=-d\sigma^+ d\sigma^- \,.
\:\)
First in the region of $x^1 >0$, $\:\rho=\rho(x^- )\,,\:$
and the equation, $\:e^{2\rho}\,dx^- =d\sigma^- \,,\:$ leads to
\begin{eqnarray}
\sigma^+ =x^+ \,, \hsi \sigma^- = x^- -\frac{1}{2\lambda}
\ln (1-\frac{M}{2\lambda} e^{2\lambda x^- } )\,.
\end{eqnarray}
The range of the new coordinates $\sigma^{\pm}$ is bounded by
\(\:
x^+ (\sigma^+ )>x^- (\sigma^- )
\:,\)
while the old coordinate is limited by
$-\infty < x^- < x_H$ with
\beql
x_H =-\,\frac{1}{2\lambda} \ln\frac{M}{2\lambda}\,.
\eeql
The inversion of the coordinate is given by
\begin{eqnarray}
x^+ = \sigma^+ \,, \hsi x^- = -\,\frac{1}{2\lambda}
\ln (\frac{M}{2\lambda} +e^{-2\lambda \sigma^- } )\,.
\end{eqnarray}
In this new coordinate $e^{-2\varphi}=e^{-2\lambda\sigma^0}$, clearly
indicating the linear dilaton vacuum.
On the other hand, the asymptotically flat coordinate for $x^1 <0$ is given by
\begin{eqnarray}
\chi^- =x^- \,, \hsi \chi^+ = x^+ -\frac{1}{2\lambda} \ln (1-\frac{M}{2\lambda}
e^{2\lambda x^+} )\,,
\end{eqnarray}
again with
$-\infty < x^+ < x_H$.

The spacetime geometry described by this metric represents the wormhole
structure. First, note that
the trajectory of the source in terms of the flat coordinate is the boundary
of the region, $x^1 >0$,
\begin{equation}
\sigma^- = \sigma^+ -\frac{1}{2\lambda}
           \ln(1-\frac{M}{2\lambda} e^{2\lambda \sigma^+} )\,.
\end{equation}
This trajectory approaches asymptotically the null line $\sigma^+ =x_H$.
Thus to an observer far away from the source
sitting at rest in the Minkowski coordinate system, the source
appears to move with acceleration.
A similar equation of the source trajectory in the left region may be written
in terms of $\chi^{\pm}$ coordinates, with the null asymptote at
$\chi^- =x_H$. Since the worldline of the source
is unique, the two forms of the trajectory must be identified.
The two coordinate patches, $(\,\sigma^+ \,,\sigma^-\,)$ and
$(\,\chi^+\,,\chi^-\,)$,
are thus joined at the source.

The Penrose diagram of this spacetime is depicted in Fig.1.
The world line of the localized source is designated by the solid arrow
at the center.
The spacetime is bounded by the lines, $x^{\pm}=\pm\,\infty$ and
$x^{\pm}=x_H$. These boundary lines are uniquely characterized by the
equation,
\(\:
(\,\nabla\,e^{-2\varphi}\,)^2 =0
\:,\)
and form the dilaton singularity, mostly harmless in much the same way
as the asymptotic region of the linear dilaton vacuum. Two portions
of the spacetime, $R$ and $L$, are disconnected causally, and are separated
by the two global event horizons, $x^{\pm}=x_H$.
For instance, an observer in the region $R$ can influence only within the
forward light cone entirely confined in $R$,
and he cannot receive any information from a person in the region $L$.
It should be clear then that this spacetime has a wormhole structure
with the global event horizon, furthermore being everywhere flat,
except being curved at the source.

We now consider the quantum effect under this wormhole background.
One loop path integral due to $N$ massless fields yields the well known
trace anomaly. Combined with the energy-momentum conservation, it determines
the form of the stress tensor components up to the two unknown
functions, $t_{\pm}(x^{\pm})$, which must be fixed by the boundary
condition of the problem at hand  \cite{cf}, \cite{cghs}.
The relevant stress tensor corresponding to the right moving flux
at $I^{-}_{L}$ is
\begin{equation}
\langle T_{x^- ,x^- } \rangle =-\,\frac{N}{12\pi}
\left(\,\partial_{-}^{2} \rho -(\partial_{-} \rho)^2 +t_- (x^- )\,
\right)\,. \label{polya}
\end{equation}
The boundary condition appropriate to the gravitational collapse is that
no incoming flux exists at the past null infinity, $x^+ =-\,\infty$,
which implies
\(\:
t_- (x^- ) =0
\:\)
since $\rho=O[e^{2\lambda\,x^+ }]$.

The energy-momentum tensor in the right half of the spacetime $R$ is then
expressed in terms of the flat $(\sigma^+ ,\sigma^- )$ coordinate as
\begin{equation}
\langle T_{\sigma^- ,\sigma^- } \rangle=\left[\frac{\partial x^-}
{\partial \sigma^-} \right]^2  \langle T_{x^- ,x^- } \rangle
=-\,\frac{N\lambda^2}{12\pi}\,
   \frac{1+\frac{4\lambda}{M} e^{-2\lambda \sigma^-} }
        {(1+\frac{2\lambda}{M} e^{-2\lambda \sigma^-} )^2 }\,.
\end{equation}
Near the horizon, $\sigma^- \rightarrow \infty$, and
\begin{equation}
\langle T_{\sigma^- ,\sigma^- } \rangle \sim -\,\frac{N\lambda^2}{12\pi}\,.
\label{hawstress}
\end{equation}
Combined with
\(\:
\langle T_{\sigma^+ ,\sigma^- } \rangle=
\langle T_{\sigma^+ ,\sigma^+ } \rangle=0\,,
\:\)
this can be understood as a thermal flux with the temperature of
$\:T=\frac{\lambda}{\pi}$;
\(\;
\rho_{thermal}=\frac{N\pi}{12} T^2
\:.\)
The precise form of the thermal spectrum for the right moving emitted
particles may be derived by the standard method of Bogoliubov transformation,
as in Ref.\cite{hw75}.
The particle emission thus derived is an indication of evaporation of the
classical wormhole.

To fully incorporate the back reaction in the semiclassical approximation,
it is imperative to analyze the semiclassical equations with the one loop
quantum correction added to the classical equations.
It is difficult to analyze this set of equations
by analytic methods for the Polyakov one loop effective action,
$\:\sqrt{-g}\,R\frac{1}{\Box}R\propto \rho\,\partial_+ \partial_-
\rho\,$,
but as a matter of principle there should be no problem to follow the
spacetime evolution.
In particular, our reversed model does not give rise to any curvature
singularity even when the quantum effect is included,
as shown in Ref.\cite{my92} by looking into the coefficient
matrix of the kinetic terms.

It is then reasonable to expect either of
the following two possibilities on the end point of Hawking radiation
to occur.
In one case the classical spacetime bounded by the dilaton
singularity is modified only at the
quantitative level by the quantum effect, and the wormhole geometry is
unchanged leaving behind the event horizon.
Since we have shown the Hawking radiation to be
independent of the mass $M$ of an arbitrary classical wormhole,
it is likely that the end
point of the evaporation in this case is a still unknown quantum remnant
of the wormhole geometry.
The information loss paradox is then resolved by the leak of
information to the disconnected world inaccessible to us;
the principle of quantum mechanics is unchanged, but the usual rule of
quantum mechanics restricted to the one side of the spacetime must
be modified.
In another case the wormhole is completely melted by the quantum
back reaction, and
the final spacetime is a flat linear dilaton vacuum.
In this case the thermal nature
of Hawking radiation is  superficial or may even not exist for a very
small mass hole,
and quantum mechanical correlation is retained as a whole.

In summary, we demonstrated in a variant of two dimensional dilaton gravity
theories that the classical end point of the collapse due to an arbitrary
localized
massive body is the wormhole. The standard one loop quantum effect induces
Hawking radiation under the wormhole background. With the quantum correction
included into the semiclassical set of equations, it is expected that
the information loss
paradox at the end point of the gravitational collapse may be resolved
in this model.

\newpage

\begin{center}
\begin{Large}

\bf{Figure Caption}

\end{Large}
\end{center}

\vs
\begin{flushleft}
\begin{large}
Figure 1
\end{large}
\end{flushleft}

The Penrose diagram of the classical wormhole spacetime.
The world line of the localized source is designated by the solid arrow
at the center.
The global event
horizon at $x^{\pm}=x_H$ separates the right and the left regions $R$ and
$L$ causally.

\end{document}